\documentclass[preprint,showpacs,superscriptaddress,preprintnumbers,amsmath,amssymb]{revtex4}
\usepackage{graphicx}
\usepackage{dcolumn}
\usepackage{bm}
\usepackage{hyperref}
\usepackage{siunitx}
\usepackage{amssymb}
\usepackage{amsthm}
\usepackage{color}

\begin{document}

\title{Magneto-transport study of top- and back-gated LaAlO$_3$/SrTiO$_3$ heterostructures}

\author{W. Liu}
\email{W.Liu@unige.ch}
\address{Department of Quantum Matter Physics, University of Geneva, 24 Quai Ernest-Ansermet, 1211 Geneva, Switzerland}
\author{S. Gariglio}
\affiliation{Department of Quantum Matter Physics, University of Geneva, 24 Quai Ernest-Ansermet, 1211 Geneva, Switzerland}
\author{A. F$\hat{\text{e}}$te}
\affiliation{Department of Quantum Matter Physics, University of Geneva, 24 Quai Ernest-Ansermet, 1211 Geneva, Switzerland}
\author{D. Li}
\affiliation{Department of Quantum Matter Physics, University of Geneva, 24 Quai Ernest-Ansermet, 1211 Geneva, Switzerland}
\author{M. Boselli}
\affiliation{Department of Quantum Matter Physics, University of Geneva, 24 Quai Ernest-Ansermet, 1211 Geneva, Switzerland}
\author{D. Stornaiuolo}
\thanks{Present address: Department of Physics, University of Naples Federico II and CNR-SPIN, Napoli, Italy}
\affiliation{Department of Quantum Matter Physics, University of Geneva, 24 Quai Ernest-Ansermet, 1211 Geneva, Switzerland}
\author{J.-M. Triscone}
\affiliation{Department of Quantum Matter Physics, University of Geneva, 24 Quai Ernest-Ansermet, 1211 Geneva, Switzerland}
\date{\today}

\begin{abstract}
We report a detailed analysis of magneto-transport properties of top- and back-gated LaAlO$_3$/SrTiO$_3$ heterostructures. Efficient modulation in magneto-resistance, carrier density, and mobility  of the two-dimensional electron liquid present at the interface is achieved by sweeping top and  back gate voltages. Analyzing those changes with respect to  the carrier density tuning, we observe that the back gate strongly modifies the electron mobility while the top gate mainly varies the carrier density. The evolution of the spin-orbit interaction is also followed as a function of top and back gating.
\end{abstract}

\maketitle
The two-dimensional electron liquid (2DEL) present at the interface between the insulating oxides LaAlO$_3$ (LAO) and SrTiO$_3$ (STO) exhibits several fascinating  properties, including superconductivity and a large spin-orbit coupling.\cite{mannhart2010science}  It has also attracted much attention in the context of device applications following   the realization of  field-effect transistors and devices with  nanoscale  dimensions.\cite{Zubko2011,Forg2012,cen2009oxide,Stornaiuolo2012,Stornaiuolo14PRB}


The large electric permittivity of the STO substrate, especially  at low temperatures,  facilitates tuning of the 2DEL properties by an electric field  effect  in  back gate transistors.\cite{Zubko2011} This back gate configuration  allows considerable control of  multiple parameters of the 2DEL such as the superconducting critical temperature,  spin-orbit interaction, carrier density, and mobility.\cite{Caviglia08, Bell2009a, CavigliaWAL, BenShalom2010a,Joshua2012,rakhmilevitch2013anomalous} Additionally,  it might simultaneously change  the confinement of the 2DEL.\cite{Bell2009a,minohara2014potential,biscaras2014scirep} This confinement modifies the bulk STO electronic structure and leads to an orbital ordering with light $d_{xy}$ and heavy $d_{xz}$/$d_{yz}$ sub-bands. \cite{Popovic2008,Salluzzo2009} Recently, there has been growing interest in employing top electrodes in  gating experiments of LAO/STO devices,\cite{li2011very,bi2014room, hosoda2013transistor,Eerkers,bal2014evidence,goswami2015nanoscale} as LAO has a large band gap and reasonably large dielectric constant.\cite{edge2006electrical}  The  transistor operation  of a top-gated LAO/STO device was   demonstrated with an on-off switch of the conductivity using less than 1\,V.\cite{Forg2012}  However, the responses  of the 2DEL to  top and back gate can be different: a first comparison revealed, indeed, a different modulation of the mobility by the two approaches.\cite{hosoda2013transistor} 


\begin{figure}[htbp]
 \begin{center}
 \includegraphics[width=\columnwidth]{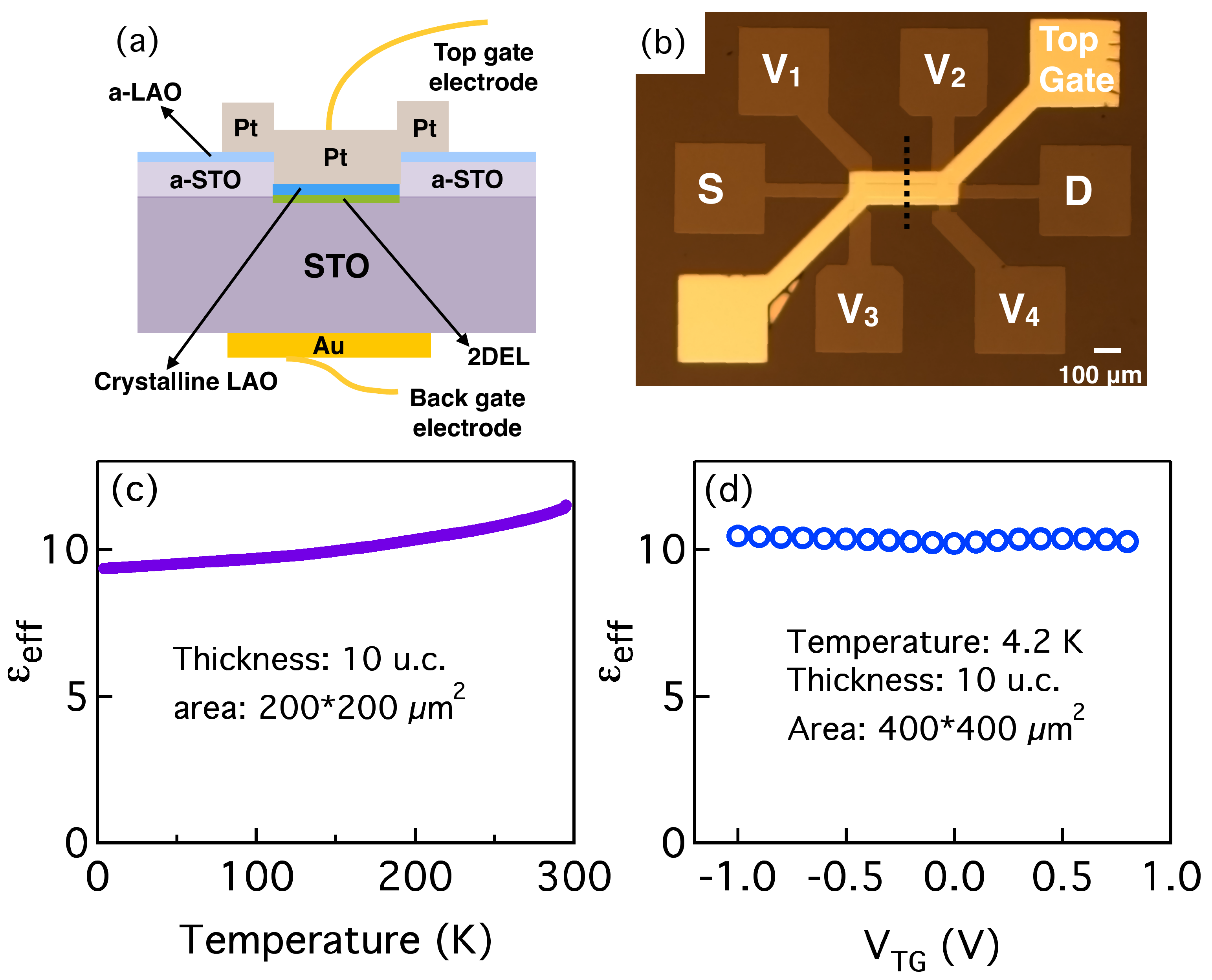}
 \caption{(a) Schematic   of the cross-section of a field-effect device with  top and back gate configuration.   ``a-LAO''  and ``a-STO'' refer to amorphous layers of LAO and STO. The 2DEL is only formed between the crystalline LAO and the STO substrate. (b) Optical image of a sample surface with a Hall bar structure with channel width  50\,$\mu$m.  Source  and drain are labeled as S and D. The  distance between the longitudinal voltage contacts  is 300\,$\mu$m. The top-gate electrode consists of  sputtered Pt film and appears golden under the optical microscope.  The black dashed line indicates the location of the cross-section  displayed  in (a). (c) and (d) Dependence of LAO dielectric constant $\epsilon_{eff}$ on temperature and   top gate voltage $V_{TG}$  at 4.2\,K  of a 10 unit cell sample. }
 \label{Figure1}
 \end{center}
 \end{figure}

In this paper, we present a systematic study of  transport  properties of top- and back-gated LAO/STO heterostructures in the presence of a  perpendicular magnetic field.  Electric field effect tuning is achieved by using the LAO film and STO substrate as top and back gate dielectrics respectively. Large tunability in resistance, carrier density and magneto-resistance were observed in both configurations.  The top-gate approach   shows reversible tuning with voltage sweeping, contrary to back gate.\cite{Caviglia08, biscaras2014scirep} Top and  back gates  affect  the mobility differently, back gate being more effective to boost it. The evolution of the weak-localization/weak-antilocalization behavior is also extracted from the magneto-transport data. This allows us to study inelastic and spin-orbit magnetic fields for constant carrier densities achieved by a combined use of top and bottom gates. We try to link the differences  in the response of the 2DEL to the confining potential shape  and to the multi-band conduction.

The Hall bars  are defined by  pre-depositing an amorphous STO layer as a hard mask.\cite{Stornaiuolo2012}   Structures down to 25\,$\mu$m  in width are realized by optical lithography. Subsequently, epitaxial  LAO films  are  deposited on TiO$_2$-terminated STO (001)-oriented single crystals  by pulsed laser deposition in an oxygen pressure of \SI{8e-5}\,Torr  at 800$^{\circ}\mathrm{C}$ or 830$^{\circ}\mathrm{C}$ and annealed in 0.2\,bar of oxygen at 520$^{\circ}\mathrm{C}$ for an hour as documented elsewhere.\cite{Caviglia08,Cancellieri2010}  After growth, \textit{ex-situ} sputtered Pt  is used for the top gate and  Au  films are sputtered  on the backside of the substrate  for the back gate  electrodes.   A schematic  of   a device with a top and back gate  is displayed in Figure \ref{Figure1}(a). Figure \ref{Figure1}(b) shows an optical image of a typical Hall bar structure with  sputtered Pt film as the top gate. Measurements are  carried out keeping the  leakage current from the top electrode below 1\,nA, much smaller than the channel current  (typically in the range 0.1 $-$ 1\,$\mu$A). Transport data presented here below were obtained measuring a sample grown at 830$^{\circ}\mathrm{C}$ with 15 unit cells (u.c.) thick crystalline LAO  in a perpendicular magnetic  field configuration. Samples with thickness in the range 10 $-$ 17\,u.c. and  channel width in the range 25 $-$ 100\,$\mu$m were also fabricated and tested:  they all show comparable  transport properties.


\begin{figure}[t]
 \begin{center}
 \includegraphics[width=\columnwidth]{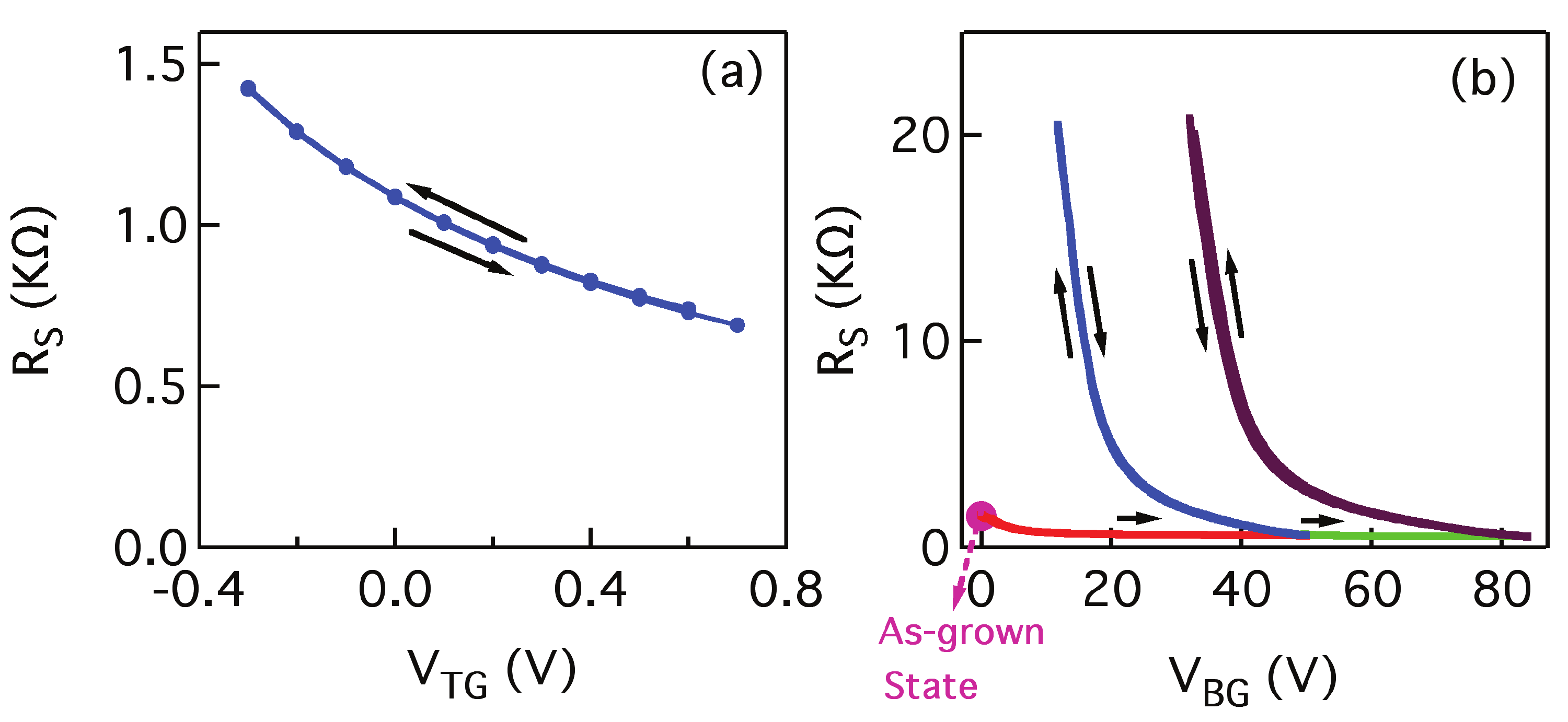}
 \caption{Sheet resistance R$_{\text{S}}$ as a function of applied $V_{TG}$  (a)  and  $V_{BG}$ (b) at 1.5\,K. A large path dependent behavior is observable upon the application of   $V_{BG}$.   See text for more details. }
 \label{Figure2}
 \end{center}
 \end{figure}
 
Capacitance   characterization  with an AC voltage of 10\,mV at 200\,Hz   demonstrates good capacitor behavior (small loss tangent and negligible leakage current) of  top-gated devices prepared in this fashion.\cite{vacuumbox}   We consistently observe an  effective dielectric constant ($\epsilon_{eff}$) for the Pt/LAO/2DEL capacitor smaller than the bulk value, which is possibly due to  interfacial effects.\cite{stengel2006origin} Moreover, this interfacial effect might  be the reason for the weak dependence of $\epsilon_{eff}$  on  temperature   as shown in Figure \ref{Figure1}(c)  (bulk LAO has a  temperature independent dielectric constant).  $\epsilon_{eff}$ has little dependence on the  top gate voltage $V_{TG}$ as shown in Figure \ref{Figure1}(d). We also do not observe in the investigated $V_{BG}$ range substantial changes  in the capacitance when we modulate the 2DEL carrier density using back gate.\cite{capacitanceofLAO}

\begin{figure}[b]
 \begin{center}
 \includegraphics[width=0.8\columnwidth]{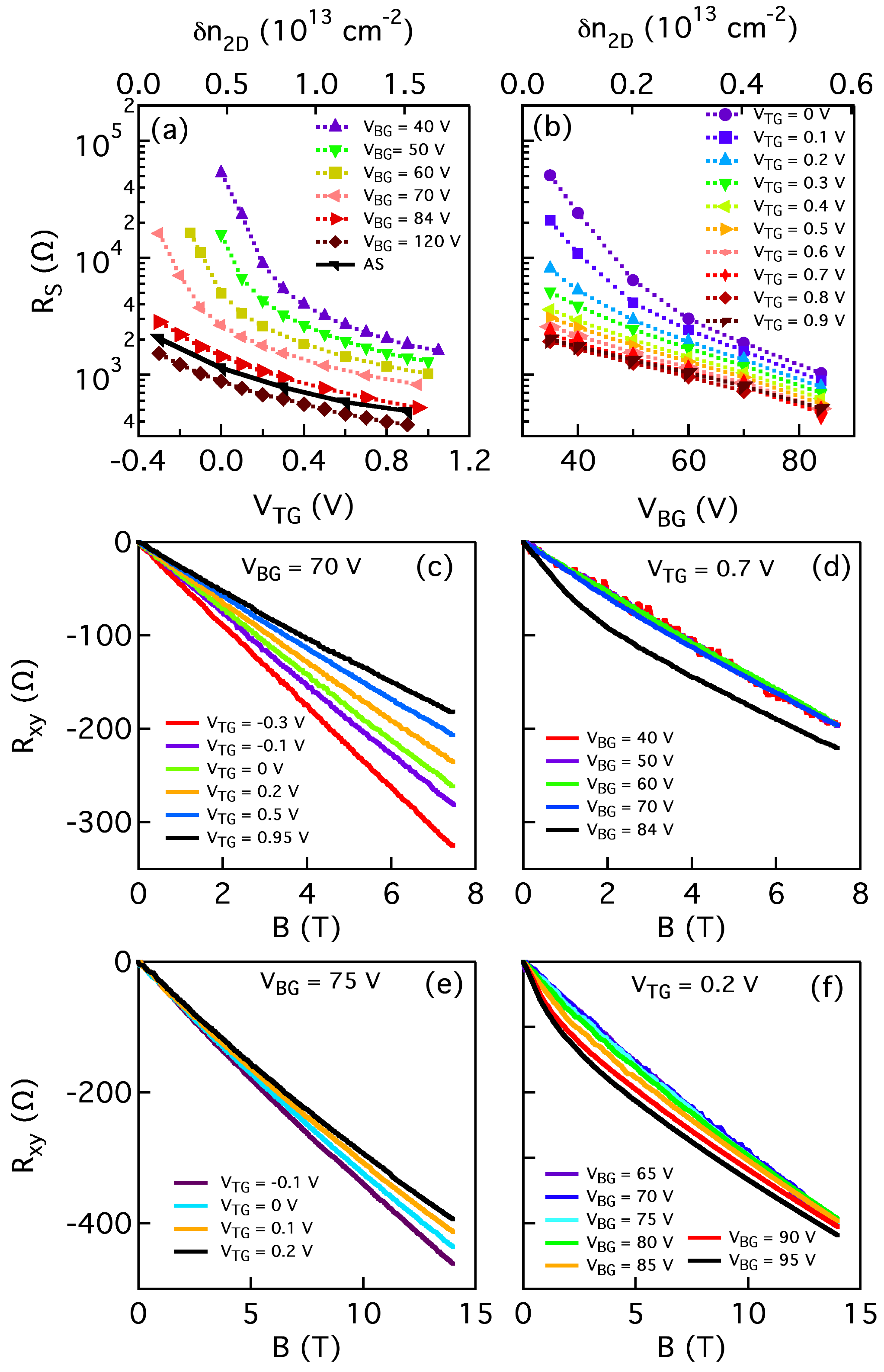}
 \caption{(a) and (b) The  change in $R_{S}$  of a 15\,u.c. thick sample is plotted either as a function of $V_{TG}$ with  a  fixed $V_{BG}$ or as a function of  $V_{BG}$ with a  fixed $V_{TG}$.  ``AS'' is short for ``As-grown State''. (c) and (d) Magnetic field dependence of the antisymmetrized $R_{xy}$ for  $V_{BG}$ = 70\,V and $V_{TG}$ = 0.7\,V. (e) and (f) Magnetic field dependence of the antisymmetrized $R_{xy}$ for  $V_{BG}$ = 75\,V and $V_{TG}$ = 0.2\,V at 700\,mK.}
 \label{Figure3}
 \end{center}
 \end{figure}

Figure \ref{Figure2} compares the low temperature resistance behavior sweeping the gate voltages from the top (panel (a)) and from the back (panel (b)).\cite{sampledetails} For top gate voltages $V_{TG}$, the tuning of the sheet resistance $R_S$ is reversible.  We note the modulation of $R_{S}$ is about 40\%, the $V_{TG}$ range being set to avoid  leakage across the LAO layer. For back gate voltages $V_{BG}$, a large history dependent behavior manifests in $R_S$ when $V_{BG}$ is first swept upwards and then downwards. In this geometry, from the as-grown state,  $R_S$ first decreases and then saturates for a wide range of $V_{BG}$ (red curve in panel (b)). If we stop increasing $V_{BG}$ (at $V^{max}_{BG}$  = 50\,V as in panel (b)) and reverse the sweeping direction, we observe that the resistance increases strongly (blue curve), deviating from the upward sweep. Then, as long as $V_{BG}$ is kept below $V^{max}_{BG}$, the sweeps are reversible (the resistance moves along the blue curve). In this configuration,  extremely high resistance states can be achieved.  If  $V_{BG}$  exceeds $V^{max}_{BG}$, the resistance remains constant in the upward sweep (green curve); however, the new reversible behavior is shifted to higher voltages (purple curve).  This ``forming process'' of the resistance state suggests that many added electrons spill out of the quantum well and go into  trapped states.\cite{biscaras2014scirep}

To build a more comprehensive  understanding  of  the different effects of   $V_{TG}$ and   $V_{BG}$, we  investigate the   magneto-transport properties of the 2DEL using both gates. The data are displayed in Figure \ref{Figure3}: in the left panels, the measurements are performed at various  $V_{TG}$ while $V_{BG}$  is maintained  fixed, and vice versa for the right panels.  We observe that the changes in $R_{S}$ induced by sweeping   $V_{TG}$ or $V_{BG}$ are quite comparable, but on a completely different voltage range. Figures \ref{Figure3}(c) - \ref{Figure3}(f) display the Hall resistance  $R_{xy}$ as a function of magnetic field for different values of $V_{TG}$  and  $V_{BG}$: we observe an evolution from linear to non-linear behavior pointing to a transition from single to multiple band conduction.\cite{BenShalom2010b,Joshua2012,Fete2012} Since correctly extracting   the  total carrier density $n_{2D}$ from a non-linear $R_{xy}$ is generally difficult, the modulation of  the  total carrier density  $\delta n_{2D}$ is calculated using a parallel plate capacitor model. For  the top gate, using  $\epsilon_{eff}$  = 12, we obtain a change in $n_{2D}$ that  agrees with the estimation from the variation of the Hall coefficient with $V_{TG}$ when $R_{xy}$ is linear in field (see Figure \ref{Figure3}(c)). For the back gate,   $\delta n_{2D}$ is obtained by the capacitor measurement as a function of $V_{BG}$.\cite{Caviglia08} These $\delta n_{2D}$ modulations are plotted as top axes in panels (a) and (b). Now comparing  the change in $R_{S}$ versus $\delta n_{2D}$ for the two gates, we see that the same  $R_{S}$ modulation is achieved by top gate with a variation in $n_{2D}$ that is roughly twice that obtained with  the back gate.

This different response of $R_{S}$ to the top and back gates, as well as the evolution of the Hall effect, may suggest  distinct effects of the two gates on the confining potential and hence on the sub-band structure. Self-consistent calculations using a Schr$\ddot{\text{o}}$dinger-Poisson approach show that a pure increase in carrier density results in a stronger confinement.\cite{Copie2009,khalsa2012theory} Modeling of  back gate transistors \cite{Bell2009a} and  analysis of back gate voltage sweeps \cite{biscaras2014scirep} indicate that positive voltages increase the confinement width (weaker confinement) of the 2DEL; as a consequence, the energy splitting between the lower energy  $d_{xy}$ - and lower mobility - and higher energy $d_{xz}$/$d_{yz}$ - and higher mobility - bands  should be reduced. 
 
\begin{figure}[t]
\begin{center}
\includegraphics[width=0.8\columnwidth]{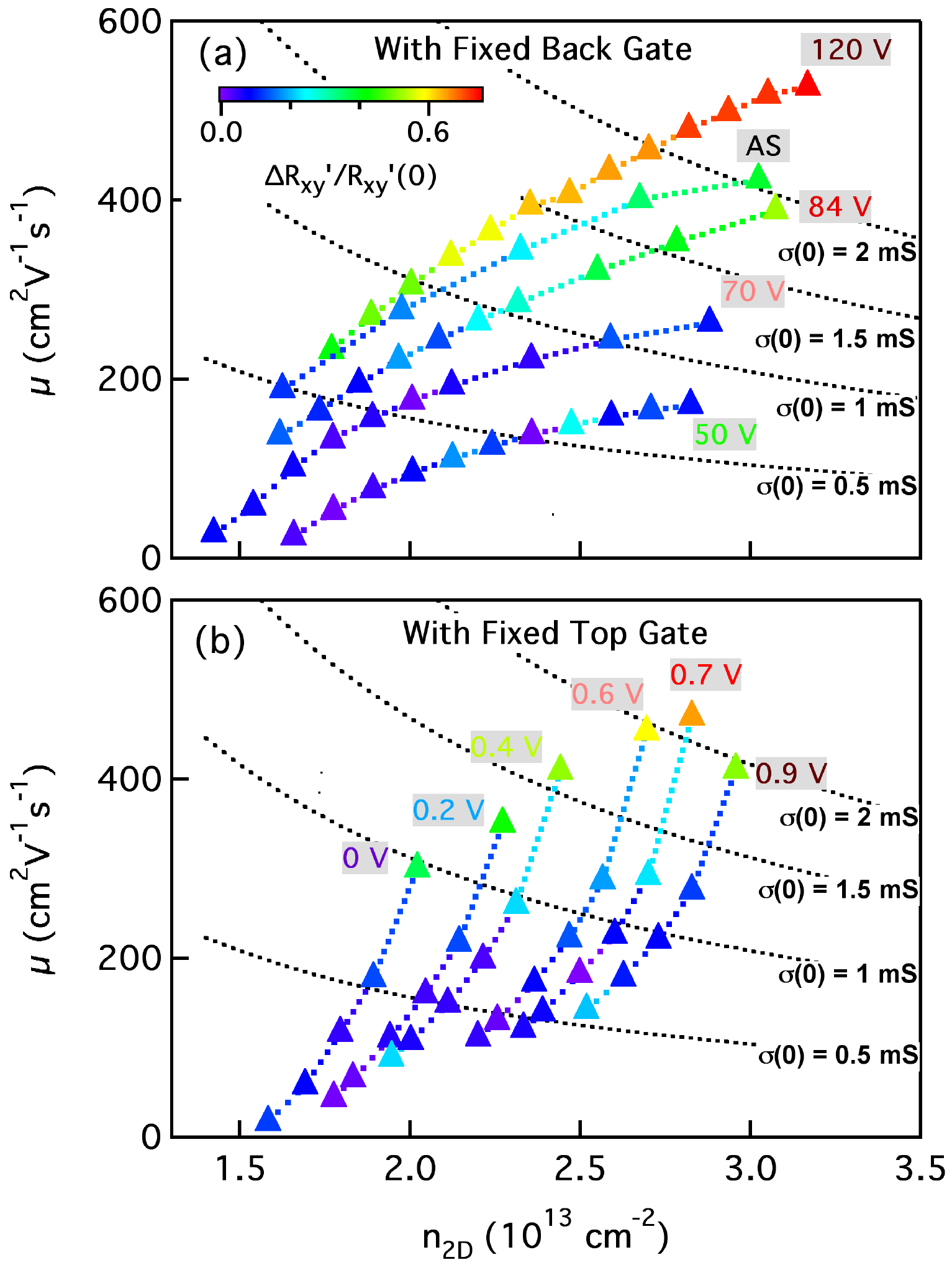}
\caption{$n_{2D}$  dependence of  the  mobility  at different fixed  $V_{BG}$ (a) and different fixed $V_{TG}$ (b). Color indicates the magnitude of  calculated $\Delta R'_{xy}/R'_{xy}$(0). (a) and (b) have the same color legend for the non-linearity.  The values of the fixed  $V_{BG}$ or $V_{TG}$ are labeled next to their corresponding curves with the same color code as in Figures \ref{Figure3}(a) and  \ref{Figure3}(b) for that particular voltage value.  For clarity, a selection of data is presented. Dashed lines in both graphs mark the contour of a constant $\sigma(0)$.}
 \label{Figure4}
 \end{center}
 \end{figure}

In order to  highlight the difference between the two gate approaches,  we plot the mobility $\mu$ as a function of the carrier density in Figure \ref{Figure4} for top (panel (a)) and back (panel (b)) gate voltage sweeps. $n_{2D}$ is estimated from the linear Hall effect and the carrier change induced by the capacitor effect,   and $\mu$ is defined as  $\sigma(0)/n_{2D}e$, where $e$ is the electron charge and  $\sigma(0)$ is the zero-field conductance; $\mu$ calculated in this fashion is therefore an effective mobility. In the same graphs, we use a color code to indicate the degree of  non-linearity observed in the Hall resistance $\Delta R'_{xy}/R'_{xy}(0) =|R'_{xy}\text{(7.5\,T)}-R'_{xy}\text{(0\,T)}|/R'_{xy}\text{(0\,T)}$, $R'_{xy}$ being the derivative $\partial R/\partial B$ at a given field. For single band conduction, $\Delta R'_{xy}/R'_{xy}(0)$ is zero (represented as blue) while for multi-band conduction, it becomes non-zero (represented as yellow/red). The mobility curves in Figure \ref{Figure4}(a) show that top gate sweeps result in a strong modulation of $n_{2D}$ and a smaller change in $\mu$. When we change the back gate voltage, we notice a sharp increase in the carrier mobility: this is more evident in panel (b) where mobility is strongly enhanced by back gate voltages for a small variation in the carrier density. We also observe that, concomitant with the increase in $\mu$, the Hall resistance becomes non-linear, suggesting that a second band with high mobility comes into play. \cite{Popovic2008,Salluzzo2009,BenShalom2010b,Joshua2012,Fete2012}
 
A possible scenario to understand the response of the transport properties to the field-effect tuning  relates the effect of the two gates to the confining potential. As discussed above, a deconfinement of the 2DEL would reduce the $d_{xy}$ versus $d_{xz}$/$d_{yz}$ band splitting, this energy becoming  negative in bulk STO. \cite{AllenPRB2013} Along with this effect, carriers could be transferred to the more mobile band, resulting in higher mobility and non-linear Hall resistance. On the other hand, if electrons are more confined, the energy splitting should increase, and a shift in the Fermi level would mainly induce an increase in $n_{2D}$ with little change in $\mu$.
 
\begin{figure}[t]
\begin{center}
\includegraphics[width=\columnwidth]{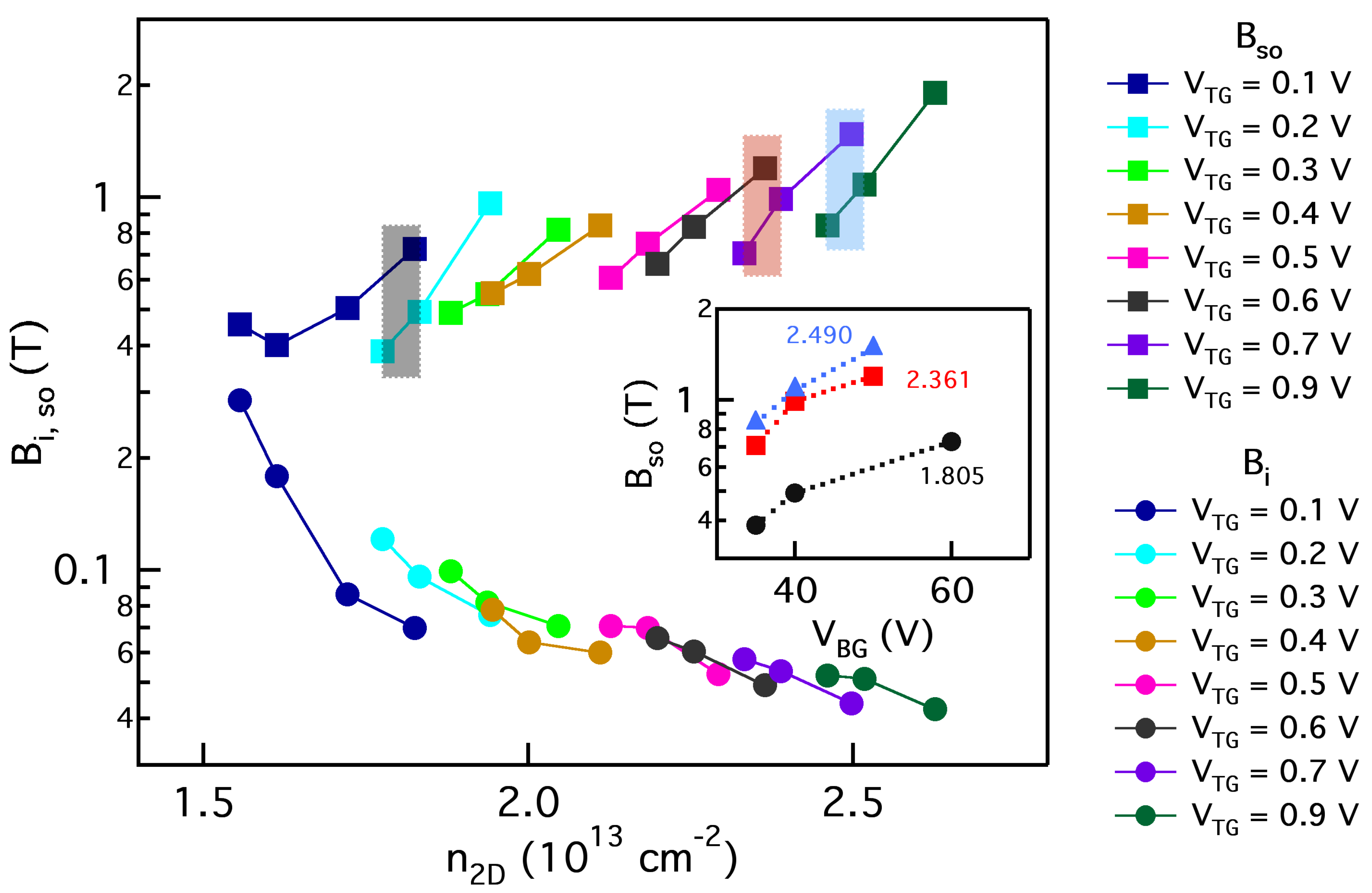}
\caption{Evolution of the inelastic $B_{i}$ and spin-orbit $B_{so}$ characteristic fields as a function of the 2D carrier density. For each $V_{TG}$ value, $V_{BG}$ was swept typically between 35 and 60\,V (see Supplementary Information). The inset shows the evolution of  $B_{so}$ as a function of $V_{BG}$  at three fixed carrier densities obtained with several combinations of $V_{TG}$ and $V_{BG}$. These three carrier densities are indicated by the  shaded areas in the main panel and are labeled next to their corresponding curve in units of 10$^{13}$\,cm$^{-2}$. The range of carrier densities explored around the mean value indicated is about $\pm$ 0.028\,10$^{13}$\,cm$^{-2}$.}
\label{Figure5}
\end{center}
\end{figure}

We finally explore the effect of top and back gates on the spin-orbit interaction of the system. We fit the magneto-conductance curves using the Maekawa-Fukuyama formula \cite{Maekawa1981} and extract the inelastic field $B_i$ and the spin-orbit field $B_{so}$ for different sets of $V_{TG}$ and $V_{BG}$.\cite{SI}  Figure \ref{Figure5} shows the evolution of $B_{i}$ and $B_{so}$ as a function of $n_{2D}$. We observe that for each $V_{TG}$ the back-gate tuning results in an effective modulation of $B_{so}$, in agreement with  previous studies on the spin-orbit interaction.\cite{CavigliaWAL,Fete2012} While $B_i$ seems to be determined mainly by $n_{2D}$, $B_{so}$ should be   more sensitive to  the modification of the confining potential. To get further insight into the evolution  of $B_{so}$  upon gate tuning, in the inset of Figure \ref{Figure5}, we plot $B_{so}$ versus $V_{BG}$ for the three carrier densities indicated by the  shaded areas in the main panel of the figure. 
For the plot, $V_{BG}$ is chosen as the ``tuning parameter" as it should affect  the electron confining potential most and hence the strength of the spin-orbit interaction, although probably  in a complex way.\cite{Joshua2012,Zhong2013,Kim2013,Khalsa2013} The data plotted in the inset show that the spin-orbit coupling can be tuned at constant $n_{2D}$ and   suggest that $V_{BG}$ is indeed modifying the confinement of the 2DEL.

To conclude, we present a systematic study  of the effects of top and back gate  on LAO/STO heterostructures.  The  detailed characterization reveals   differences between the evolution of $R_{S}$, $R_{xy}$ and  magneto-conductance using the two approaches.  The observation for the back gate of a forming process   and the strong enhancement of the electron mobility suggest that positive back gate deconfines the 2DEL: on one side some electrons are trapped deep in the STO; on the other side, the contribution from $d_{xz}$/$d_{yz}$  states becomes more significant. The control of the electronic properties of the 2DEL using both gates, as demonstrated here, allows the spin-orbit strength to be tuned at fixed carrier densities and opens new opportunities to realize transistors where mobility and carrier density are tuned independently.

We acknowledge Pavlo Zubko for useful experimental advice and Marc Gabay for enlightening discussions. We  thank M. Lopes and S. C. M$\ddot{\text{u}}$ller for the technical assistance. This research was  supported by the Swiss National Science Foundation through the
NCCR MaNEP and Division II and  has received funding from the European Research Council under the European Unionʼs Seventh Framework
Programme (FP7/2007–2013) / ERC Grant Agreement no. 319286 (Q-MAC). 

 \clearpage
 \newpage

\section{SUPPLEMENTARY MATERIAL}
\begin{figure}[t]
\includegraphics[width=\columnwidth,angle=0]{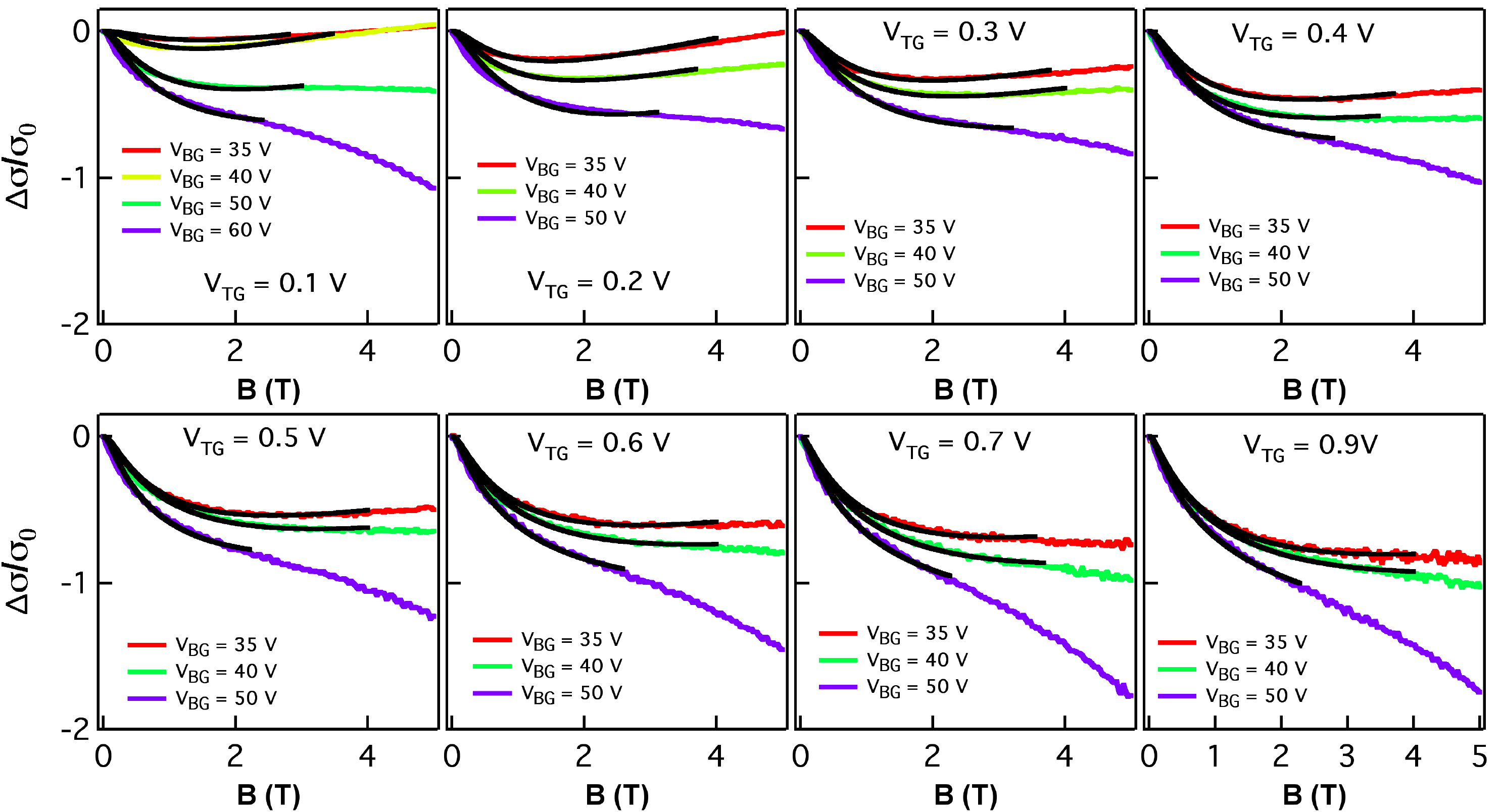}
\caption{Magneto-conductance curves for back gate voltage sweeps for selected different fixed top gates. The black solid lines in all the panels are the fits performed using MF formula.}
\label{SIFig2}
\end{figure}
In this supplementary information, we  fit the  magneto-conductance curves using the Maekawa-Fukuyama (MF) formula:\cite{Maekawa1981,CavigliaWAL} 
\begin{eqnarray*}
\Delta \sigma(B)/\sigma_0 &=& \Psi\left(\frac{B}{B_i+B_{so}}\right)
						+\frac{1}{2\sqrt{1-\gamma^2}}\Psi\left(\frac{B}{B_i+B_{so}(1+\sqrt{1-\gamma^2})}\right)\\
&&-\frac{1}{2\sqrt{1-\gamma^2}}\Psi\left(\frac{B}{B_i+B_{so}(1-\sqrt{1-\gamma^2})}\right),
\end{eqnarray*}
where $\Psi(x) = ln(x) + \psi(\frac{1}{2}+\frac{1}{x})$ with $\psi(x)$ the digamma function, and $\gamma = g\mu_BB/4eDB_{so}$. Here $\sigma(B) = \frac{R_S(B)}{R^2_S(B)+R^2_{xy}(B)}$ and $\sigma_0 = e^2/\pi h$ is a universal value of conductance.

Figure \ref{SIFig2} displays magneto-conductance curves.  In each panel, the top gate is fixed and the back gate is varied.  For low doping samples, the data seem to reflect the evolution of the weak-localization/weak-antilocalization behavior  as reported in references \citenum{CavigliaWAL} and \citenum{BenShalom2010a}. At high dopings and high magnetic fields, data deviate from weak anti-localization fittings, making the estimation of the fitting parameters less reliable.\cite{yuan2013zeeman}


\begin{thebibliography}{39}
\bibitem[{vac()}]{vacuumbox}
\bibinfo{note}{We note that keeping the samples in a vacuum environment after
  lithography and before Pt sputtering improves the dielectric response of the
  top-gated heterostructures.}
\bibitem[{sam()}]{sampledetails}
\bibinfo{note}{Data were obtained from different patterns of the same sample.}
\end{thebibliography}
\end{document}